# Interference Relay Channels – Part I: Transmission Rates

Brice Djeumou, *Student Member, IEEE,* Elena Veronica Belmega, *Student Member, IEEE,* and Samson Lasaulce, *Member, IEEE*

**Abstract**

We analyze the performance of a system composed of two interfering point-to-point links where the transmitters can exploit a common relay to improve their individual transmission rate. When the relay uses the amplify-and-forward protocol we prove that it is not always optimal (in some sense defined later on) to exploit all the relay transmit power and derive the corresponding optimal amplification factor. For the case of the decode-and-forward protocol, already investigated in [1], we show that this protocol, through the cooperation degree between each transmitter and the relay, is the only one that naturally introduces a game between the transmitters. For the estimate-and-forward protocol, we derive two rate regions for the general case of discrete interference relay channels (IRCs) and specialize these results to obtain the Gaussian case; these regions correspond to two compression schemes at the relay, having different resolution levels. These schemes are compared analytically in some special cases. All the results mentioned are illustrated by simulations, given in this part, and exploited to study power allocation games in multi-band IRCs in the second part of this two-part paper.

## I. INTRODUCTION

The case of unlicensed frequency bands has become more and more important over the last decade in part because of the resounding success of Wi-Fi systems. In these bands, wireless devices generally communicate

B. Djeumou, E. V. Belmega and S. Lasaulce are with LSS (joint lab of CNRS, Supélec, Univ. Paris Sud 11), Supélec, Plateau du Moulon, 91192 Gif-sur-Yvette, France, {djeumou,belmega,lasaulce}@lss.supelec.fr.

in a non-coordinated manner which therefore leads to a scenario where signals interfere. In order to limit interference, devices have to transmit with a relatively (e.g., in comparison with a cellular phone) low power ($17-20$ dBm typically) which implies, in particular, some limitations on the communication range and rate. A quite natural way of improving the range, transmission rate or/and quality of the communications is to add relaying nodes that can be exploited by certain/all devices operating in the same frequency band. The mentioned scenario therefore gives one (but not the unique) strong motivation for studying the following system: a network comprising two transmitters communicating with their respective receivers and a relaying node that can be used by both transmitters. In this two-part paper, we analyze two key aspects of this type of networks, which is modeled by an interference relay channel (IRC) as defined in [1][2] and re-defined later on. In the first part, we focus on the relaying protocols for such a system. Our goal is to know how the three dominant classes of protocols, namely amplify-and-forward (AF), decode-and-forward (DF) and estimate-and-forward (EF), can be adapted, used and compared in terms of transmission rate in the context of interest. In the second part, we assume that the two transmitter-receiver pairs can communicate over several bands for which a relay is available on each of them (parallel IRC). The considered framework corresponds to the powerful paradigm of cognitive wireless networks [3][4] where transmitters can sense their environment and react accordingly. More specifically, in our context, the transmitters have to decide by themselves the best power allocation (PA) policy between the available bands in order to maximize their transmission rates. In order to analyze how the transmitters compete for the additional spectral resources and relaying nodes, we will exploit the results derived in the first part of the paper and also game theoretical tools which will be used to model the interaction between the transmitters and prove the existence of predictable and stable states (in the sense of Nash [5]) for these networks.

In this part of the paper, one of our goals is to assess the performance of the network under study when terminals implement good channel codes. This is why we will consider Shannon transmission rates as the performance measure. Note that, even in real systems where such good codes are not always implemented, considering Shannon rates can bring some insights on how to design the system. Indeed, there is a direct relationship between the achievable transmission rate of a user and its signal-to-interference plus noise ratio (SINR). Therefore, optimizing Shannon rates can also allow one to optimize performance metrics like the SINR or related quantities of the same type (e.g., the carrier-to-interference ratio). As far as the interference relay channel is concerned, note that the general idea of introducing a relay in a multiuser channel is quite natural and therefore has been exploited by other authors for other types of networks. For example, the authors of [6] introduced the multiple access relay channel for which two transmitters can exploit a relay to communicate with a common receiver. The authors of [7][8] studied the dual of this problem, the broadcast relay channel where a single transmitter can cooperate with a relay to communicate with two receivers. As for the IRC itself, it has been introduced by [1][2]. In [1] and [2], the authors assumed that the relay implements the DF protocol and a (non-causal) dirty paper coding (DPC) scheme respectively, and derived achievable transmission rate regions for Gaussian IRCs. Part I of the present paper precisely aims at providing new results for the channel introduced by [1][2]. As mentioned in the abstract, our contribution is essentially fourfold: 1) we derive an achievable transmission rate region for Gaussian IRCs when the relay implements a zero-delay scalar AF (ZDSAF) protocol and prove that there is an optimum amplification factor which does not always correspond to saturating the transmit power constraint at the relay (Sec. III ); 2) assuming the DF protocol, we prove the existence of a game in Gaussian IRCs where the strategy of a transmitter is merely the correlation degree between its signal and the cooperative signal sent by the relay (Sec. IV); 3) we derive two achievable rate regions in the general case of discrete IRCs, corresponding to two EF-based protocols (Sec. V) using respectively one and two resolution levels (the Gaussian case readily follows); 4) we further discuss and compare all of these protocols by simulations (Sec. VI). Before tackling these issues, we first present, in the following section (Sec. II), the system model under consideration and main corresponding notations.

## II. System model and main notations

Fig. 1 depicts the system model considered in the whole paper except for the case of discrete input discrete output IRCs which will be studied in Sec. V-A. The system comprises two transmitters, two receivers and a relaying node. The transmitters or information source nodes are denoted by $\mathcal{S}_1$ and $\mathcal{S}_2$, the relay node by $\mathcal{R}$ and the two receivers or destination nodes by $\mathcal{D}_1$ and $\mathcal{D}_2$. As indicated in Fig. 1, the corresponding transmitted and received signals are respectively denoted by $X_1$, $X_2$, $X_r$, $Y_1$, $Y_2$ and $Y_r$. The signal $X_1$ (resp. $X_2$) conveys the private message $W_1$ (resp. $W_2$) that $\mathcal{S}_1$ (resp. $\mathcal{S}_2$) sends to $\mathcal{D}_1$ (resp. $\mathcal{D}_2$). The general form of the received baseband signals is as follows:

$$\begin{cases} Y_1 &= h_{11}X_1 + h_{21}X_2 + h_{r1}X_r + Z_1 \\ Y_2 &= h_{22}X_2 + h_{12}X_1 + h_{r2}X_r + Z_2 \\ Y_r &= h_{1r}X_1 + h_{2r}X_2 + Z_r \end{cases} \quad (1)$$

where: $\forall (i,j) \in \{1,2,r\}^2$, $h_{ij}$ represents the channel gain of the link between nodes $i$ and $j$, with the convention $h_{rr} = 0$; $Z_1$, $Z_2$ and $Z_r$ are additive white complex Gaussian noises with variances $N_1$, $N_2$ and $N_r$, respectively; the transmitted signals $X_1$, $X_2$, $X_r$ are subject to power constraints $\mathbb{E}|X_1|^2 \leq P_1$, $\mathbb{E}|X_2|^2 \leq P_2$ and $\mathbb{E}|X_r|^2 \leq P_r$. Note that, in Eq. (1), we dot not use any time index. This choice is made not only for the sake of clarity (time index is in fact required for proving the provided theorems) but also to indicate that we assume the propagation delays to be negligible. Always for the sake of clarity and simplicity, the channel gains between the different nodes (see Fig. 1) will be considered fixed for the whole duration of the transmission. Physically speaking, this means that the assumed model can at least account for the path loss effects. In fact, by using standard arguments (see e.g., [9] and [10]), the presented results can be readily re-used for other types of channels like fast fading channels. Note that the expression of the signal transmitted by the relay (i.e., $X_r$) depends on the protocol assumed and will therefore be provided in each of the three sections corresponding to the three types of protocols considered. What is general, however, is that the relay is assumed to operate in the full-duplex mode. Here again, this assumption has been well discussed in the related literature [11][12] and will not prevent us from deriving results that are also applicable to the half-duplex mode, just as those obtained in other papers like [13].

*Specific notations.* Throughout this two-part paper we will use the following notations. The capacity function for complex signals is denoted by $C(x) \triangleq \log_2(1+x)$. For any real $a \in [0,1]$, the quantity $\overline{a}$ will stand for $\overline{a} = 1 - a$. The notation $-i$ will mean that $-i = 1$ if $i = 2$ and $-i = 2$ if $i = 1$. For any complex number, $c \in \mathbb{C}$, $c^*$, $|c|$ and $\mathrm{Re}(c)$ will stand for the complex conjugate, absolute value and real part, respectively.

## III. TRANSMISSION RATES FOR THE ZDSAF PROTOCOL

In this section, the relay is assumed to implement an analog amplifier which does not introduce any delay on the relayed signal. The signal transmitted by the relay merely writes as $X_r = a_r Y_r$ where $a_r$ corresponds to the relay amplification factor/gain. We will always assume $a_r$ to be fixed. In the described setup, the following theorem provides a region of transmission rates that can be achieved when the transmitters send private messages to their respective receivers, the relay implements the ZDSAF protocol and the receivers implement single-user decoding (i.e., no multiuser detection or successive interference cancellation schemes are allowed here).

*Theorem 3.1 (Transmission rate region for the IRC with ZDSAF):* Let $R_i$, $i \in \{1,2\}$, be the Shannon

*transmission rate for the source node $\mathcal{S}_i$. When ZDSAF is assumed the following region is achievable:*

$$\forall i \in \{1,2\}, \; R_i \leq C\left(\frac{|a_r h_{ir} h_{ri} + h_{ii}|^2 \rho_i}{|a_r h_{jr} h_{ri} + h_{ji}|^2 \rho_j \frac{N_j}{N_i} + a_r^2 |h_{ri}|^2 \frac{N_r}{N_i} + 1}\right) \quad (2)$$

where $\rho_i = \frac{P_i}{N_i}$ and $j = -i$.

The proof of this theorem is standard [14] and will therefore be omitted. The main point to be mentioned is that Gaussian codebooks have to be assumed to obtain the proposed region. An interesting point to investigate is the choice of the value of the amplification gain $a_r$. In the vast majority of the papers available in the literature, $a_r$ is chosen in order to saturate the power constraint at the relay ($\mathbb{E}|X_r|^2 = P_r$) that is: $a_r = \overline{a}_r = \sqrt{\frac{P_r}{\mathbb{E}|Y_r|^2}} = \sqrt{\frac{P_r}{|h_{1r}|^2 P_1 + |h_{2r}|^2 P_2 + N_r}}$. As mentioned in some works [15][16][17][18], this choice can turn out to be sub-optimal in the sense of certain performance criteria. The intuitive reason for this is that the AF protocol not only amplifies the useful signal but also the received noise. While this effect can be negligible in certain scenarios for the standard relay channel, it is generally a dominant effect for the IRC. Indeed, even if the noise at the relay is negligible, the interference term for user $i$ (i.e., the term $h_{jr} X_j$, $j = -i$) is generally not. This gives us a particular motivation for choosing the amplification factor $a_r$ adequately that is, to maximize the transmission rate of a given user or the network sum-rate. The proposed derivation differs from [15][17] because here we consider a different system (an IRC instead of a relay channel with no direct link), a specific relaying function (linear relaying functions instead of arbitrary functions) and a different performance metric (individual transmission rate and sum-rate instead of raw bit error rate [15] and mutual information [17]). Our problem is also different from [18] since we do not consider the optimal clipping threshold in the sense of the end-to-end distortion for frequency division relay channels. At last, the main difference with [16] is that, for the relay channel, the authors discuss the choice of the optimal amplification gain in terms of transmission rate for a vector AF protocol having a delay of at least one symbol duration; here we focus on a scalar AF protocol with no delay and a different system namely the IRC. In this setup, we have found an analytical expression for the best $a_r$ in the sense of $R_i(a_r)$ for a given user $i \in \{1, 2\}$ and seen that the $a_r$ maximizing the network sum-rate has to be computed numerically in general. The corresponding analytical result is stated in the following theorem.

*Theorem 3.2:* [Optimal amplification gain for the ZDSAF in the IRC] *The transmission rate of user $i$, $R_i(a_r)$, as a function of $a_r \in [0, \overline{a}_r]$ can have several critical points which are the real solutions, denoted* by $a_{r,i}^{(1)}$ and $a_{r,i}^{(2)}$, to the following equation:

$$a_r^2 \left[|m_i|^2 \mathrm{Re}(p_i q_i^*) - (|p_i|^2 + s_i)\mathrm{Re}(m_i n_i^*)\right] + a_r \left[|m_i|^2(|q_i|^2 + 1) - |n_i|^2(|p_i|^2 + s_i)\right] \\ + (|q_i|^2 + 1)\mathrm{Re}(m_i n_i^*) - |n_i|^2 \mathrm{Re}(p_i q_i^*) = 0 \quad (3)$$

where $m_i = h_{ir} h_{ri} \sqrt{\rho_i}$, $n_i = h_{ii} \sqrt{\rho_i}$, $p_i = h_{jr} h_{ri} \sqrt{\rho_j}$, $q_i = h_{ji} \sqrt{\rho_j}$, $s_i = |h_{ri}|^2$, $i \in \{1,2\}$ and $j = -i$. Thus, depending on the channel parameters, the optimal amplification gain $a_r^* = \arg \max_{a_r \in [0, \overline{a}_r]} R_i(a_r)$ takes a value in the set $a_r^* \in \{0, \overline{a}_r, a_{r,i}^{(1)}, a_{r,i}^{(2)}\}$. If, additionally, the channel gains are reals then the two critical points write as: $a_{r,i}^{(1)} = -\frac{n_i}{m_i}$ and $a_{r,i}^{(2)} = -\frac{m_i q_i^2 + m_i - p_i q_i n_i}{m_i q_i p_i - p_i^2 n_i - n_i s_i}$.

The proof of this result is provided in Appendix A. Of course, in practice, if the receive SINR (viewed from a given user) at the relay is low, choosing the amplification factor $a_r$ adequately does not solve the problem. It is well known that in real systems, a more efficient way to combat noise is to implement error correcting codes. This is one of the reasons why DF is also an important relaying protocol, especially for digital relay transceivers for which AF cannot be implemented in its standard form (see e.g., [18] for more details).

## IV. Transmission rates for the DF protocol

The purpose of this section is essentially twofold. First, we state a corollary from [1]. Indeed, the given result corresponds to the special case of the rate region derived in [1] where each source sends to its respective destination a private message only (and not both public and private messages as in [1]). The corresponding rate region is provided for making this paper sufficiently self-containing and for being used in the simulation part to establish a comparison between the different relaying protocols under consideration in this paper. Second, we explain why the DF protocol naturally introduces a game between the transmitters. The proof of the existence of Nash equilibrium in the corresponding game will be provided in Part II. The principle of the DF protocol is detailed in [13] and here we just give the main idea behind it. Consider a Gaussian relay channel for which the source-relay link has a better quality than the source-destination link. From each message intended for the destination, the source builds a coarse and a fine message. With these two messages, the source superposes two codewords. The rates associated with these codewords (or messages) are such that the relay can decode both of them reliably while the destination can only decode the coarse message. After decoding this message, the destination can subtract the corresponding signal and try to decode the fine message. To help the destination to do so, the relay cooperates with the source by sending some information about the fine message. Mathematically, this explanation translates in the IRC as follows. The signal transmitted by $\mathcal{S}_i$ is structured as $X_i = X_{i0} + \sqrt{\frac{\tau_i}{\nu_i}\frac{P_i}{P_r}} X_{ri}$ where: the signals $X_{i0}$ and $X_{ri}$ are independent and precisely correspond to the coarse and fine messages respectively; the parameter $\nu_i$ represents the fraction of transmit power the relay allocates to user $i$, hence we have $\nu_1 + \nu_2 \leq 1$; the parameter $\tau_i$ represents the fraction of transmit power $\mathcal{S}_i$ allocates to the cooperation signal. Using these notations we have the following result.

*Corollary 4.1 ([1]): When DF is assumed, the following region is achievable; for $i \in \{1,2\}$,*

$$R_i \leq \min\left\{C\left(\frac{|h_{ir}|^2(1-\tau_i)P_i}{N_r}\right), C\left(\frac{|h_{ii}|^2 P_i + |h_{ri}|^2 \nu_1 P_r + 2\mathcal{R}e(h_{ii}h_{ri}^*)\sqrt{\tau_i P_i \nu_i P_r}}{|h_{ji}|^2 P_j + |h_{ri}|^2 \nu_j P_r + 2\mathcal{R}e(h_{ji}h_{ri}^*)\sqrt{\tau_j P_j \nu_j P_r} + N_i}\right)\right\} \quad (4)$$

*where $j = -i$, $(\nu_1, \nu_2) \in [0,1]^2$ s.t. $\nu_1 + \nu_2 \leq 1$ and $(\tau_1, \tau_2) \in [0,1]^2$.*

In a context of decentralized networks or in the case of unlicensed bands, each source $\mathcal{S}_i$ has to optimize the parameter $\tau_i$ in order to maximize its transmission rate $R_i$. As shown in the above rate region, this choice is not independent of the choice of the other source. Therefore, each source finds its optimal strategy by optimizing its rate w.r.t. $\tau_i^*(\tau_j)$. In order to do that, each source has to make some assumptions on the value $\tau_j$ used by the other source. This is precisely a non-cooperative game where each player makes some assumptions on the other player's behavior and maximizes its own utility. We therefore see that the DF protocol, through the parameter $\tau_i$ representing the cooperation degree between the source $\mathcal{S}_i$ and the relay, introduces a PA game. In fact, one can even derive a hierarchical game since the relay can also be thought of as a player tuning $(\nu_1, \nu_2)$ (i.e., the fractions of power dedicated to users $1$ and $2$). All these issues will be treated properly in Part II.

## V. Transmission rates for the EF protocol

In this section, we consider a third main class of relaying protocols, namely the estimate-and-forward protocol. The EF protocol can be implemented in digital transceivers (contrarily to the standard AF protocol) and always allows the receiver(s) to improve its (their) performance with respect to the non-cooperative case (in contrast with DF protocols that can degrade the performance in practice, see e.g., [18]). In order to derive the corresponding transmission rate region for the Gaussian IRC, we have to prove that this region is effectively achievable. For this purpose, we first tackle the general case of discrete input discrete output channels. We know, from standard quantization and continuity arguments [14], that the Gaussian case can be readily obtained from the discrete case. The proofs based on Shannon codes are not necessary for a good understanding and interpretation of the derived results and are therefore detailed in Appendices B and

C. Before providing these results, we will first review the principle of the EF protocol which allows the reader to better understand the structure of this section. The principle of the EF protocol in the standard relay channel is that the relay sends an approximated version of its observation signal to the receiver. More precisely, in its information-theoretic standard version [13], it consists in compressing the received signal at the relay in the Wyner-Ziv manner [19] i.e., knowing that the destination also receives a direct signal from the source that is correlated with the signal to be compressed. The compression rate is precisely tuned by taking into account this correlation degree and the quality of the relay-destination link. In our setup i.e., an IRC with two receivers, the EF protocol can be designed in at least two manners. On the one hand, the relay can define two resolution levels for its observation signal, which is what we call bi-level compression. On the other hand, it can also use a single resolution level adapted to the worse destination but reliably decodable by the better one (we call this scheme single-level compression). One of our contributions is to derive the transmission rates that can be achieved by using these two schemes and then discuss the difference between them.

## A. The case of discrete IRCs

Before providing the two theorems associated with the two EF protocols proposed, we first define the discrete IRC.

*Definition 5.1:* A two-user discrete memoryless interference relay channel (DMIRC) without feedback consists of three input alphabets $\mathcal{X}_1$, $\mathcal{X}_2$ and $\mathcal{X}_r$, and three output alphabets $\mathcal{Y}_1$, $\mathcal{Y}_2$ and $\mathcal{Y}_r$, and a probability transition function that satisfies $p(y_1^n, y_2^n, y_r^n \mid x_1^n, x_2^n, x_r^n) = \prod_{k=1}^{n} p(y_{1,k}, y_{2,k}, y_{r,k} \mid x_{1,k}, x_{2,k}, x_{r,k})$ for some $n \in \mathbb{N}^*$.

For the EF protocol based on a bi-level compression, the relay constructs two estimated versions of its observed signal $Y_r$, which are denoted by $\hat{Y}_{r1}$ and $\hat{Y}_{r2}$. As shown in Appendix C, these estimates are constructed on a block-by-block basis: the relay constructs, from a block of $n$ observations $y_r^n = (y_r(1), ..., y_r(n))$ two codewords $\hat{y}_{r1}^n$ and $\hat{y}_{r2}^n$ intended for $\mathcal{D}_1$ and $\mathcal{D}_2$ respectively. The signal transmitted by the relay results from mixing these two codewords. To this end, the coding scheme we used is superposition coding, which explains the presence of the coding auxiliary variables $U_1$, $U_2$ in the theorem below; the meaning of these variables will be made obvious in the Gaussian case.

*Theorem 5.2:* For the DMIRC $(\mathcal{X}_1 \times \mathcal{X}_2 \times \mathcal{X}_r, p(y_1, y_2, y_r | x_1, x_2, x_r), \mathcal{Y}_1 \times \mathcal{Y}_2 \times \mathcal{Y}_r)$ with private messages and bi-level compression EF protocol, any transmission rate pair $(R_1, R_2)$ is achievable, where

$$R_1 \leq I(X_1; Y_1, \hat{Y}_{r1}|U_1) \text{ and } R_2 \leq I(X_2; Y_2, \hat{Y}_{r2}|U_2), \tag{5}$$

*under the constraints*

$$I(Y_r; \hat{Y}_{r1}|U_1, Y_1) \leq I(U_1; Y_1) \text{ and } I(Y_r; \hat{Y}_{r2}|U_2, Y_2) \leq I(U_2; Y_2), \tag{6}$$

*for some joint distribution*

$$p(x_1, x_2, u_1, u_2, x_r, y_1, y_2, y_r, y_{r1}, y_{r2}, \hat{y}_{r1}, \hat{y}_{r2}) = $$
$$p(x_1)p(x_2)p(u_1)p(u_2)p(x_r|u_1, u_2)p(y_1, y_2, y_r|x_1, x_2, x_r)p(\hat{y}_{r1}|y_r, u_1)p(\hat{y}_{r2}|y_r, u_2).$$

In the case of single-level compression, the relay compresses $Y_r$ and encodes the same estimate $\hat{Y}_r$ for both destination nodes. A natural question to be asked is whether the single-level resolution case is a special case of the bi-level case. The answer is no. Although the relaying principle is the same in both cases, as the Appendices show, there is one technical difference in the encoding procedure. In the bi-level case, the codewords used for superposition coding (associated with $U_1$ and $U_2$) at the relay are independent by construction. In the single-level case, there is a unique codeword for both destination. This is why making the choice $U_1 \equiv U_2 \equiv X_r$ in the bi-level case to obtain the single-level case is not admissible since the corresponding codewords would be totally correlated whereas they have to be independent. Technically, the choice $U_1 \equiv U_2 \equiv X_r$ would violate the Markov chains assumed in Theorem 5.2. Rather, the case of single-level compression can be stated through the following theorem.

*Theorem 5.3:* For the DMIRC $(\mathcal{X}_1 \times \mathcal{X}_2 \times \mathcal{X}_r, p(y_1, y_2, y_r|x_1, x_2, x_r), \mathcal{Y}_1 \times \mathcal{Y}_2 \times \mathcal{Y}_r)$ with private messages and single-level compression EF protocol, any transmission rate pair $(R_1, R_2)$ is achievable, where

$$R_i \leq I\left(X_i; Y_i, \hat{Y}_r \mid X_r\right), \quad i \in \{1, 2\}, \tag{7}$$

*under the constraint*

$$\max_i I(Y_r; \hat{Y}_r|X_r, Y_i) \leq \min_i I(X_r; Y_i), \tag{8}$$

*for some joint distribution* $p(x_1, x_2, x_r, y_1, y_2, y_r, y_{r1}, y_{r2}, \hat{y}_r) = p(x_1)p(x_2)p(x_r)p(y_1, y_2, y_r|x_1, x_2, x_r)p(\hat{y}_r|y_r, x_r).$

## B. The case of Gaussian IRCs

As mentioned in the beginning of this section, obtaining achievable transmission rates for Gaussian IRCs from those for discrete IRCs is an easy task. Indeed, the latter consists in using Gaussian codebooks everywhere and choosing the coding auxiliary variables properly (i.e., $U_1$, $U_2$, $\hat{Y}_{r,1}$ and $\hat{Y}_{r,2}$ in the bi-level case and, $X_r$ and $\hat{Y}_r$ in the single-level case).

For the bi-level compression, the coding auxiliary variables $U_1$ and $U_2$ are chosen to be independent and distributed as $U_1 \sim \mathcal{N}(0, \nu_1 P_r)$ and $U_2 \sim \mathcal{N}(0, \nu_2 P_r)$. The corresponding codewords $u_1^n$ and $u_2^n$ convey the messages resulting from the compression of $Y_r$. The auxiliary variables $\hat{Y}_{r,1}$, $\hat{Y}_{r,2}$ write as $\hat{Y}_{r,1} = Y_r + Z_{wz}^{(1)}$ and $\hat{Y}_{r,2} = Y_r + Z_{wz}^{(2)}$ where the compression noises $Z_{wz}^{(1)} \sim \mathcal{N}(0, N_{wz}^{(1)})$ and $Z_{wz}^{(2)} \sim \mathcal{N}(0, N_{wz}^{(2)})$ are independent. At last, the relay transmits the signal $X_r = U_1 + U_2$ as in the case of a broadcast channel except that, here, each destination also receives two direct signals from the source nodes. By making these choices of random variables in Theorem 5.2 we obtain the following theorem.

*Theorem 5.4:* For the Gaussian IRC with private messages and bi-level compression EF protocol, any rate pair $(R_1, R_2)$ is achievable where

1) if $C\left(\frac{|h_{r1}|^2 \nu_2 P_r}{|h_{11}|^2 P_1 + |h_{21}|^2 P_2 + |h_{r1}|^2 \nu_1 P_r + N_1}\right) \geq C\left(\frac{|h_{r2}|^2 \nu_2 P_r}{|h_{22}|^2 P_2 + |h_{12}|^2 P_1 + |h_{r2}|^2 \nu_1 P_r + N_2}\right)$, we have

$$R_1 \leq C\left(\frac{|h_{11}|^2 P_1}{N_1 + \frac{|h_{21}|^2 P_2 \left(N_r + N_{wz}^{(1)}\right)}{|h_{2r}|^2 P_2 + N_r + N_{wz}^{(1)}}} + \frac{|h_{1r}|^2 P_1}{N_r + N_{wz}^{(1)} + \frac{|h_{2r}|^2 P_2 N_1}{|h_{21}|^2 P_2 + N_1}}\right), \quad (9)$$

$$R_2 \leq C\left(\frac{|h_{22}|^2 P_2}{N_2 + |h_{r2}|^2 \nu_1 P_r + \frac{|h_{12}|^2 P_1 \left(N_r + N_{wz}^{(2)}\right)}{|h_{1r}|^2 P_1 + N_r + N_{wz}^{(2)}}} + \frac{|h_{2r}|^2 P_2}{N_r + N_{wz}^{(2)} + \frac{|h_{1r}|^2 P_1(|h_{r2}|^2 \nu_1 P_r + N_2)}{|h_{12}|^2 P_1 + |h_{r2}|^2 \nu_1 P_r + N_2}}\right), \quad (10)$$

subject to the constraints $N_{wz}^{(1)} \geq \frac{\left(|h_{11}|^2 P_1 + |h_{21}|^2 P_2 + N_1\right) A - A_1^2}{|h_{r1}|^2 \nu_1 P_r}$ and $N_{wz}^{(2)} \geq \frac{\left(|h_{22}|^2 P_2 + |h_{12}|^2 P_1 + |h_{r2}|^2 \nu_1 P_r + N_2\right) A - A_2^2}{|h_{r2}|^2 \nu_2 P_r}$,

2) else, if $C\left(\frac{|h_{r2}|^2 \nu_1 P_r}{|h_{22}|^2 P_2 + |h_{12}|^2 P_1 + |h_{r2}|^2 \nu_2 P_r + N_2}\right) \geq C\left(\frac{|h_{r1}|^2 \nu_1 P_r}{|h_{11}|^2 P_1 + |h_{21}|^2 P_1 + |h_{r1}|^2 \nu_2 P_r + N_1}\right)$, we have

$$R_1 \leq C\left(\frac{|h_{11}|^2 P_1}{N_1 + |h_{r1}|^2 \nu_2 P_r + \frac{|h_{21}|^2 P_2 \left(N_r + N_{wz}^{(1)}\right)}{|h_{2r}|^2 P_2 + N_r + N_{wz}^{(1)}}} + \frac{|h_{1r}|^2 P_1}{N_r + N_{wz}^{(1)} + \frac{|h_{2r}|^2 P_2(|h_{r1}|^2 \nu_2 P_r + N_1)}{|h_{21}|^2 P_2 + |h_{r1}|^2 \nu_2 P_r + N_1}}\right), \quad (11)$$

$$R_2 \leq C\left(\frac{|h_{22}|^2 P_2}{N_2 + \frac{|h_{12}|^2 P_1 \left(N_r + N_{wz}^{(2)}\right)}{|h_{1r}|^2 P_1 + N_r + N_{wz}^{(2)}}} + \frac{|h_{2r}|^2 P_2}{N_r + N_{wz}^{(2)} + \frac{|h_{1r}|^2 P_1 N_2}{|h_{12}|^2 P_1 + N_2}}\right), \quad (12)$$

subject to the constraints $N_{wz}^{(1)} \geq \frac{\left(|h_{11}|^2 P_1 + |h_{21}|^2 P_2 + |h_{r1}|^2 \nu_2 P_r + N_1\right) A - A_1^2}{|h_{r1}|^2 \nu_1 P_r}$ and $N_{wz}^{(2)} \geq \frac{\left(|h_{22}|^2 P_2 + |h_{12}|^2 P_1 + N_2\right) A - A_2^2}{|h_{r2}|^2 \nu_2 P_r}$,

3) else

$$R_1 \leq C\left(\frac{|h_{11}|^2 P_1}{N_1 + |h_{r1}|^2 \nu_2 P_r + \frac{|h_{21}|^2 P_2 \left(N_r + N_{wz}^{(1)}\right)}{|h_{2r}|^2 P_2 + N_r + N_{wz}^{(1)}}} + \frac{|h_{1r}|^2 P_1}{N_r + N_{wz}^{(1)} + \frac{|h_{2r}|^2 P_2(|h_{r1}|^2 \nu_2 P_r + N_1)}{|h_{21}|^2 P_2 + |h_{r1}|^2 \nu_2 P_r + N_1}}\right), \quad (13)$$

$$R_2 \leq C\left(\frac{|h_{22}|^2 P_2}{N_2 + |h_{r2}|^2 \nu_1 P_r + \frac{|h_{12}|^2 P_1 \left(N_r + N_{wz}^{(2)}\right)}{|h_{1r}|^2 P_1 + N_r + N_{wz}^{(2)}}} + \frac{|h_{2r}|^2 P_2}{N_r + N_{wz}^{(2)} + \frac{|h_{1r}|^2 P_1(|h_{r2}|^2 \nu_1 P_r + N_2)}{|h_{12}|^2 P_1 + |h_{r2}|^2 \nu_1 P_r + N_2}}\right), \quad (14)$$

*subject to the constraints* $N_{wz}^{(1)} \geq \frac{\left(|h_{11}|^2 P_1 + |h_{21}|^2 P_2 + |h_{r1}|^2 \nu_2 P_r + N_1\right)A - A_1^2}{|h_{r1}|^2 \nu_1 P_r}$ *and* $N_{wz}^{(2)} \geq \frac{\left(|h_{22}|^2 P_2 + |h_{12}|^2 P_1 + |h_{r2}|^2 \nu_1 P_r + N_2\right)A - A_2^2}{|h_{r2}|^2 \nu_2 P_r}$, *with* $(\nu_1, \nu_2) \in [0,1]^2, \nu_1 + \nu_2 \leq 1$, $A = |h_{1r}|^2 P_1 + |h_{2r}|^2 P_2 + N_r$, $A_1 = 2\mathcal{R}e(h_{11} h_{1r}^*) P_1 + 2\mathcal{R}e(h_{21} h_{2r}^*) P_2$ *and* $A_2 = 2\mathcal{R}e(h_{12} h_{1r}^*) P_1 + 2\mathcal{R}e(h_{22} h_{2r}^*) P_2$.

The three scenarios emphasized in this theorem corresponds to the following three situations: 1) $\mathcal{D}_1$ has the better link (in the sense of the theorem) and can decode both the relay message intended for $\mathcal{D}_2$ and its own message and can therefore cancel the interference signal due to the relay; 2) this scenario is the dual of scenario 1); 3) in this latter scenario, none of the destination is able to suppress the interference generated by the relay.

For the single-level compression, one just needs to choose $\hat{Y}_r$ as $\hat{Y}_r = Y_r + Z_{wz}$ with $Z_{wz} \sim \mathcal{N}(0, N_{wz})$ and apply Theorem 5.3. We find the following result.

*Theorem 5.5:* For the Gaussian IRC with private messages and single-level compression EF protocol, the rate pair $(R_1, R_2)$ is achievable, where

$$R_i \leq C\left(\frac{|h_{ii}|^2 P_i}{N_i + \frac{|h_{ji}|^2 P_j (N_r + N_{wz})}{|h_{jr}|^2 P_j + N_r + N_{wz}}} + \frac{|h_{ir}|^2 P_i}{N_r + N_{wz} + \frac{|h_{jr}|^2 P_j N_j}{|h_{ji}|^2 P_j + N_i}}\right) \quad (15)$$

*subject to the constraint* $N_{wz} \geq \frac{\max\{\sigma_1^2, \sigma_2^2\}}{2^{2R_0} - 1}$, *with*

$$R_0 = \min\left\{C\left(\frac{|h_{r1}|^2 P_r}{|h_{11}|^2 P_1 + |h_{21}|^2 P_2 + N_1}\right), C\left(\frac{|h_{r2}|^2 P_r}{|h_{22}|^2 P_2 + |h_{12}|^2 P_1 + N_2}\right)\right\}, \quad (16)$$

*and* $\sigma_i^2 = |h_{ir}|^2 P_i + |h_{jr}|^2 P_j + N_r - \frac{\left(2\mathcal{R}e(h_{ii} h_{ir}^*) P_i + 2\mathcal{R}e(h_{ji} h_{jr}^*) P_j\right)^2}{|h_{ii}|^2 P_i + |h_{ji}|^2 P_i + N_i}$, *with* $i \in \{1, 2\}$ *and* $j = -i$.

With this type of EF protocols, the relay adapts the compression resolution to the worse receiver. At this point, one interesting question arises. Is it better to choose the bi-level compression EF protocol or the single-level compression EF protocol? This is one of the issues treated in the next section, which is dedicated to simulations. It turns out that, in some useful special cases, an analytical answer can be given. This is the purpose of the end of this section.

The first special case is the case of asymmetric channels. By asymmetric we mean that $N_2 \gg N_1$ or $N_1 \gg N_2$. Consider w.l.o.g. the case $N_2 \gg N_1$ with $N_2 \to +\infty$ and $N_1 < +\infty$. It is obvious and easy to check that $R_2 \to 0$ for both the single-level (SL) and bi-level (BL) compression schemes. For the rate of

user 1 we have:

$$\lim_{N_2 \to \infty} R_1^{(\text{SL})} = C\left(\frac{|h_{11}|^2 P_1}{|h_{21}|^2 P_2 + N_1}\right) \tag{17}$$

$$\lim_{N_2 \to \infty} R_1^{(\text{BL})} = C\left(\frac{|h_{11}|^2 P_1}{N_1 + \frac{|h_{21}|^2 P_2 \left(N_r + N_{wz}^{(1)}\right)}{|h_{2r}|^2 P_2 + N_r + N_{wz}^{(1)}}} + \frac{|h_{1r}|^2 P_1}{N_r + N_{wz}^{(1)} + \frac{|h_{2r}|^2 P_2 N_1}{|h_{21}|^2 P_2 + N_1}}\right). \tag{18}$$

It is clear that user 1 gets a higher rate with the bi-level compression scheme and, as the rate for user 2 tends to zero, this protocol is also the better one in terms of network sum-rate. A second case, which is very special but has the advantage of requiring no long derivations is the perfectly symmetric IRC, which we define by $P_1 = P_2$, $N_1 = N_2$, $|h_{1r}| = |h_{2r}|$, $|h_{12}| = |h_{21}|$, $|h_{11}| = |h_{22}|$ and $|h_{r1}| = |h_{r2}|$. With this setting, a quick inspection of the compression noise expressions in Theorems 5.2 and 5.3 shows that the compression noise level (i.e., $N_{wz}$) of the single-level scheme will always be lower than those (i.e., $N_{wz}^{(1)}$ and $N_{wz}^{(2)}$) obtained with the bi-level scheme. This is because, in symmetric IRCs, in addition to the fact that both receivers can perform almost similarly in terms of wyner-ziv compression noise level, the single-level compression scheme does not generate additional interference whereas, in the case of bi-level compression, there is always at least one user undergoing some additional interference from the relay.

## VI. SIMULATION RESULTS

*Simulation setup.* In all this section we assess the performance of the IRC in terms of system sum-rate as a function of the relay position. For this purpose we assume a path loss model for the channel gains $|h_{ij}|$ and a given location for each node. For the path loss model we take $|h_{ij}| = \left(\frac{d_{ij}}{d_0}\right)^{-\frac{\gamma}{2}}$ for $(i,j) \in \{1, 2, r\}^2$ where $d_0 = 5$ m is a reference distance and $\gamma = 2$ is the path loss exponent. The nodes $\mathcal{S}_1, \mathcal{S}_2, \mathcal{D}_1, \mathcal{D}_2$ are assumed to be in a plane. The positions of the nodes will be indicated on each figure considered and are characterized in this plane by the distance between the nodes which are chosen as follows: $d'_{11} = 11.5$ m, $d'_{22} = 10$ m, $d'_{12} = 11$ m and $d'_{21} = 14$ m. As for the relay, to avoid any divergence for the path loss in $d_{ij} = 0$, we assume that it is not in this plane but at $\epsilon = 0.1$ m from it i.e., the relay location is given by the $(x_r, y_r, z_r)$ where $z_r$ is fixed and equals $0.1$ m; thus $d_{ij} = \sqrt{d'^2_{ij} + \epsilon^2}$ for $i = r$ or $j = r$ and $i \neq j$. The noise levels at the receiver nodes are assumed to be normalized and unitary ($N_1 = N_2 = N_r = 1$). In terms of transmit power we analyze two cases: a symmetric case where $P_1 = P_2 = 10$ (normalized power) and an asymmetric one where $P_1 = 3$ and $P_2 = 10$. The relay transmit power is fixed: $P_r = 10$. We have seen that, for both DF and bi-level compression EF protocols, the relay has to allocate its power between the two cooperative signals intended for the two receivers. We consider two different cases: the uniform PA policy ($\nu_1 = \nu_2 = \frac{1}{2}$) and the optimal PA policy $(\nu_1^*, \nu_2^*) \triangleq \arg\max_{(\nu_1,\nu_2)} R_1 + R_2$ s.t. $\nu_1 + \nu_2 \leq 1$ that maximizes the system sum-rate.

*Comparing the different relaying protocols.* We start by investigating the best choice for the relaying protocol between the AF, DF and bi-level compression EF protocols. For the symmetric scenario, Fig. 2 represents the regions of the plane $\left(\frac{x_r}{d_0}, \frac{y_r}{d_0}\right) \in [-4, +4] \times [-3, +4]$ (corresponding to the possible relay positions) where one given protocol performs better than the two others in terms of system sum-rate. These regions are in agreement with what is generally observed for the standard relay channel. This type of information is useful, for example, when the relay has to be in some places because of some practical constraints and one has to choose the best protocol for a given location. Fig. 3 allows one to better quantify the differences in terms of sum-rate between the AF, DF and bi-level EF protocols since it represents the sum-rate versus $x_r$ for a given $y_r = 0.5d_0$. The discontinuity observed stems from the fact that for the bi-level EF protocol there is a frontier delineating the scenarios where one receiver is better than the other and can therefore suppress the interference of the relay (as explained in Sec. V-B). We indicate this frontier in the figures commented just below.

*Single level compression vs. bi-level compression.* Here we focus on the EF protocol and its two variants proposed. Fig. 4 corresponds to the symmetric scenario and $\nu_1 = \nu_2 = \frac{1}{2}$ and shows the regions where the relay performs better by using the bi-level compression scheme instead of the single-level one. The dotted line represents, for the bi-level compression EF protocol, the frontier separating the two scenarios "receiver 1 is better than receiver 2" and conversely (in the sense of Theorem 5.4). These regions illustrate a combination of several trade-offs, which we try to explain clearly here. When the qualities of the links $\mathcal{R} \to \mathcal{D}_i$ and $\mathcal{R} \to \mathcal{D}_j$ (with $j = -i$) are quite similar and it is also the case between the links $\mathcal{S}_i \to \mathcal{R}$ and $\mathcal{S}_j \to \mathcal{R}$, the single-level scheme is the most efficient because: the compression rate for each receiver is not limited by the other receiver; the relay uses all its power $P_r$ to transmit the cooperative signal; as there is only one cooperation message, the relay generates no additional interference. This scheme is thus the best choice both for $R_1$ and $R_2$ and therefore for $R_1 + R_2$. On the other hand, if the qualities of the links $\mathcal{R} \to \mathcal{D}_i$ and $\mathcal{R} \to \mathcal{D}_j$ are markedly different, which happens when the relay is located around $\mathcal{D}_i$ (resp. $\mathcal{D}_j$), it is better for $R_i$ (resp. $R_j$) to choose the double-level scheme. This is true if the negative effect that the double-level scheme allocates only $\frac{P_r}{2}$ to each cooperation message is compensated by the positive effect that the compression noise level seen at $\mathcal{D}_i$ (resp. $\mathcal{D}_j$) will no longer be lower-bounded by the one at $\mathcal{D}_j$ (resp. $\mathcal{D}_i$). Fig. 4 precisely illustrates the complex trade-offs just mentioned. Fig. 5 shows these trade-offs for the asymmetric scenario.

## VII. CONCLUSION

In this part we have derived achievable transmission rate regions for the interference relay channel when the relay respectively implements the AF, DF and EF protocols. In comparison with the standard relay channel, we have emphasized several important differences. For the AF protocol, from a given receiver's point of view (say from $\mathcal{D}_1$), the relay not only amplifies the noise but also the interference due to $\mathcal{S}_2$. As a consequence, scenarios where the amplification factor used by the relay does not saturate the power constraint at the relay are much more frequent than for a relay channel. We have seen that the optimal factor can be determined analytically. For the DF protocol we have seen to what extent it creates a game in interference relay channels. For the EF protocol, we have seen that because of presence of several receivers, the EF protocol can be designed based on multi-level compression, which leads to several variants of the EF protocol (the single-level and bi-level compression schemes in our case of two receivers). The best choice between them depends on the performance criterion to be optimized (a given individual rate or the network sum-rate) and the network topology (symmetric/asymmetric networks). The provided theoretical and simulation results gives insights on how to choose the relaying protocol in IRCs or locate a relay optimally for a given protocol. An extension of this work is proposed in Part II where we investigate the case of multi-band IRCs with cognitive transmitters that can freely and selfishly allocate their power between the available bands.

## APPENDIX A
### OPTIMAL AMPLIFICATION GAIN AT THE RELAY WITH THE ZDSAF PROTOCOL

Using the notations given in Theorem 3.2 and also the SINR in the capacity function of Eq. (3.1) the rate $R_i$ can be written as:

$$R_i(a_r) = C\left(\frac{|m_i a_r + n_i|^2}{|p_i a_r + q_i|^2 + s_i a_r^2 + 1}\right),$$

We observe that $R_i(0) = C\left(\frac{|n_i|^2}{|q_i|^2+1}\right)$ and that we have an horizontal asymptote $R_{i,\infty} \triangleq \lim_{a_r \to \infty} R_1(a_r) = C\left(\frac{|m_i|^2}{|p_i|^2 + s_i}\right)$. Also the first derivative w.r.t. $a_r$ is

$$R'_i(a_r) = \frac{a_r^2\left[|m_i|^2\mathrm{Re}(p_i q_i^*) - (|p_i|^2 + s_i)\mathrm{Re}(m_i n_i^*)\right] + a_r\left[|m_i|^2(|q_i|^2 + 1) - |n_i|^2(|p_i|^2 + s_i)\right] + (|q_i|^2 + 1)\mathrm{Re}(m_i n_i^*) - |n_i|^2\mathrm{Re}(p_i q_i^*)}{[|p_i a_r + q_i|^2 + s_i a_r^2 + 1][|m_i a_r + n_i|^2 + |p_i a_r + q_i|^2 + s_i a_r^2 + 1]}$$

The explicit solution, $a_r^*$ depends on the channel parameters and is given here below. We denote by $\Delta$ the discriminant of the nominator in the previous equation. If $\Delta < 0$, then in function of the sign of $|m_i|^2 \mathrm{Re}(p_i q_i^*) - (|p_i|^2 + s_i)\mathrm{Re}(m_i n_i^*)$, the function $R_i(a_r)$ is either decreasing and $a_r^* = 0$ or increasing and $a_r^* = \overline{a}_r$. Let us now focus on the case where $\Delta \geq 0$.

1) If $|m_i|^2 \mathrm{Re}(p_i q_i^*) - (|p_i|^2 + s_i)\mathrm{Re}(m_i n_i^*) \geq 0$ then
    a) if $a_{r,i}^{(1)} \leq 0$ and $a_{r,i}^{(2)} \leq 0$ then $a_r^* = \overline{a}_r$;
    b) if $a_{r,i}^{(1)} > 0$ and $a_{r,i}^{(2)} \leq 0$ then
        i) if $\overline{a}_r \geq a_{r,i}^{(1)}$ then $a_r^* = 0$;
        ii) if $\overline{a}_r < a_{r,i}^{(1)}$ then
            - if $R_i(0) \geq R_i(\overline{a}_r)$ then $a_r^* = 0$ else $a_r^* = \overline{a}_r$;
    c) if $a_{r,i}^{(1)} \leq 0$ and $a_{r,i}^{(2)} > 0$ then the analysis is similar to the previous case and $a_r^* \in \{0, \overline{a}_r\}$ depending on $a_r^{(2)}$ this time;
    d) if $a_{r,i}^{(1)} > 0$ and $a_{r,i}^{(2)} > 0$
        i) if $a_{r,i}^{(1)} < a_{r,i}^{(2)}$
            A) if $\overline{a}_r \leq a_{r,i}^{(1)}$ then $a_r^* = \overline{a}_r$;
            B) if $a_{r,i}^{(1)} < \overline{a}_r \leq a_{r,i}^{(2)}$ then $a_r^* = a_{r,i}^{(1)}$;
            C) if $\overline{a}_r > a_{r,i}^{(2)}$ then
                - if $R_i(a_{r,i}^{(1)}) \geq R_1(\overline{a}_r)$ then $a_r^* = a_{r,i}^{(1)}$ else $a_r^* = \overline{a}_r$;
        ii) if $a_{r,i}^{(1)} > a_{r,i}^{(2)}$ then the analysis is similar to the previous case, exchanging the roles of $a_{r,i}^{(1)}$ and $a_{r,i}^{(2)}$;
        iii) if $a_{r,i}^{(1)} = a_{r,i}^{(2)}$ then $a_r^* = \overline{a}_r$.
2) If $|m_i|^2 \mathrm{Re}(p_i q_i^*) - (|p_i|^2 + s_i)\mathrm{Re}(m_i n_i^*) < 0$ then
    a) if $a_{r,i}^{(1)} \leq 0$ and $a_{r,i}^{(2)} \leq 0$ then $a_r^* = 0$;
    b) if $a_{r,i}^{(1)} > 0$ and $a_{r,i}^{(2)} \leq 0$ then
        i) if $\overline{a}_r \geq a_{r,i}^{(1)}$ then $a_r^* = \overline{a}_r$ else $a_r^* = a_{r,i}^{(1)}$;
    c) if $a_{r,i}^{(1)} \leq 0$ and $a_{r,i}^{(2)} > 0$ then the analysis is similar to the previous case and $a_r^* \in \{\overline{a}_r, a_{r,i}^{(1)}\}$ depending on $a_{r,i}^{(2)}$ this time;
    d) if $a_{r,i}^{(1)} > 0$ and $a_{r,i}^{(2)} > 0$
        i) if $a_{r,i}^{(1)} < a_{r,i}^{(2)}$

A) if $\overline{a}_r \leq a_{r,i}^{(1)}$ then $a_r^* = 0$;

B) if $a_{r,i}^{(1)} < \overline{a}_r \leq a_{r,i}^{(2)}$ then
   - if $R_i(0) \geq R_i(\overline{a}_r)$ then $a_r^* = 0$ else $a_r^* = \overline{a}_r$;

C) if $\overline{a}_r > a_{r,i}^{(2)}$ then
   - if $R_i(a_{r,i}^{(2)}) \geq R_i(0)$ then $a_r^* = a_{r,i}^{(2)}$ else $a_r^* = 0$;

ii) if $a_{r,i}^{(1)} > a_{r,i}^{(2)}$ then the analysis is similar to the previous case, exchanging the roles of $a_{r,i}^{(1)}$ and $a_{r,i}^{(2)}$;

iii) if $a_{r,i}^{(1)} = a_{r,i}^{(2)}$ then $a_r^* = 0$.

We observe that, depending on the channel parameters, it is not always optimal to saturate the power constraint at the relay.

## APPENDIX B
### ACHIEVABILITY PROOF OF THE RATE REGION IN THEOREM 5.2

**Definitions and notations**

We denote by $A_\epsilon^{(n)}(X)$ the weakly $\epsilon$-typical set for the random variable $X$. If $X$ is a discrete variable, $X \in \mathcal{X}$, then $\|\mathcal{X}\|$ denotes the cardinality of the finite set $\mathcal{X}$. We use $x^n$ to indicate the vector $(x_1, x_2, \ldots, x_n)$.

*Definition B.1:* A $(2^{nR_1}, 2^{nR_2}, n)$-code for the DMIRC with private messages consists of two sets of integers $\mathcal{W}_1 = \{1, \ldots, 2^{nR_1}\}$ and $\mathcal{W}_2 = \{1, \ldots, 2^{nR_2}\}$, two encoders: $f_i : \mathcal{W}_i \rightarrow \mathcal{X}_i^n$, a set of relay functions $\{f_{r,k}\}_{k=1}^n$ such that $x_{r,k} = f_{r,k}(y_{r,1}, y_{r,2}, \ldots, y_{r,k-1})$, $1 \leq k \leq n$ and two decoding functions $g_i : \mathcal{Y}_i^n \rightarrow \mathcal{W}_i$, $i \in \{1, 2\}$. The source node $\mathcal{S}_i$ intends to transmit $W_i$, the private message, to the receiver node $\mathcal{D}_i$.

*Definition B.2:* The average probability of error is defined as the probability that the decoded message pair differs from the transmitted message pair; that is, $P_e^{(n)} = \Pr[g_1(Y_1^n) \neq W_1 \text{ or } g_2(Y_2^n) \neq W_2 \mid (W_1, W_2)]$, where $(W_1, W_2)$ is assumed to be uniformly distributed over $\mathcal{W}_1 \times \mathcal{W}_2$. We also define the the average probability of error for each receiver as $P_{ei}^{(n)} = \Pr[g_i(Y_i^n) \neq W_i \mid W_i]$. We have $0 \leq \max\{P_{e1}^{(n)}, P_{e2}^{(n)}\} \leq P_e^{(n)} \leq P_{e1}^{(n)} + P_{e2}^{(n)}$. Hence $P_e^{(n)} \rightarrow 0$ implies that both $P_{e1}^{(n)} \rightarrow 0$ and $P_{e2}^{(n)} \rightarrow 0$, and conversely.

*Definition B.3:* A rate pair $(R_1, R_2)$ is said to be achievable for the IRC if there exists a sequence of $(2^{nR_1}, 2^{nR_2}, n)$ codes with $P_e^{(n)} \rightarrow 0$ as $n \rightarrow \infty$.

**Overview of coding strategy**

At the end of the block $k$, the relay constructs two estimations $\hat{y}_{r1}^n(k)$ and $\hat{y}_{r2}^n(k)$ of its observation $y_r^n(i)$

that intends to transmit to the receivers $\mathcal{D}_1$ and $\mathcal{D}_2$ to help them resolve the uncertainty on $w_{1,k}$ and $w_{2,k}$ respectively at the end of the block $k+1$.

**Details of the coding strategy**

Codebook generation

i     Generate $2^{nR_i}$ i.i.d. codewords $x_i^n(w_i) \sim \prod_{k=1}^n p(x_{i,k})$, where $w_i \in \{1, \ldots, 2^{nR_i}\}$, $i \in \{1, 2\}$.

ii     Generate $2^{nR_0^{(1)}}$ i.i.d. codewords $u_1^n \sim \prod_{k=1}^n p(u_{1,k})$. Label these $u_1^n(s_1)$, $s_1 \in \{1, \ldots, 2^{nR_0^{(1)}}\}$.

iii     Generate $2^{nR_0^{(2)}}$ i.i.d. codewords $u_2^n \sim \prod_{k=1}^n p(u_{2,k})$. Label these $u_2^n(s_2)$, $s_1 \in \{1, \ldots, 2^{nR_0^{(2)}}\}$.

iv     For each pair $(u_1^n(s_1), u_2^n(s_2))$, choose a sequence $x_r^n$ where $x_r^n \sim p(x_r^n | u_1^n(s_1), u_2^n(s_2)) = \prod_{k=1}^n p(x_{r,k} | u_{1,k}(s_1), u_{2,k}(s_2))$.

v     For each $u_1^n(s_1)$, generate $2^{n\hat{R}_1}$ conditionally i.i.d. codewords $\hat{y}_{r1}^n \sim \prod_{k=1}^n p(\hat{y}_{r1k} | u_{1,k}(s_1))$ and label them $\hat{y}_{r1}^n(z_1|s_1)$, $z_1 \in \{1, \ldots, 2^{n\hat{R}_1}\}$. For each pair $(u_1, \hat{y}_{r1}) \in \mathcal{U}_1 \times \hat{\mathcal{Y}}_{r1}$, the conditional probability $p(\hat{y}_{r1}|u_1)$ is defined as $p(\hat{y}_{r1}|u_1) = \sum_{x_1, x_2, y_1, y_2, y_r} p(x_1) p(x_2) p(y_1, y_2, y_r | x_1, x_2, x_r) p(\hat{y}_{r1} | y_r, u_1)$.

vi     For each $u_2^n(s_2)$, generate $2^{n\hat{R}_2}$ conditionally i.i.d. codewords $\hat{y}_{r2}^n \sim \prod_{k=1}^n p(\hat{y}_{r2k} | u_{2,k}(s_2))$ and label them $\hat{y}_{r2}^n(z_2|s_2)$, $z_2 \in \{1, 2^{n\hat{R}_2}\}$. For each triplet $(u_2, \hat{y}_{r1}) \in \mathcal{U}_2 \times \hat{\mathcal{Y}}_{r1}$, the conditional probability $p(\hat{y}_{r2}|u_2)$ is defined as $p(\hat{y}_{r2}|u_2) = \sum_{x_1, x_2, y_1, y_2, y_r} p(x_1) p(x_2) p(y_1, y_2, y_r | x_1, x_2) p(\hat{y}_{r2} | y_r, u_2)$.

vii     Randomly partition the message set $\{1, 2, \ldots, 2^{n\hat{R}_1}\}$ into $2^{nR_0^{(1)}}$ sets $\{S_1^{(1)}, S_2^{(1)}, \ldots, S_{2^{nR_0^{(1)}}}^{(1)}\}$ by independently and uniformly assigning each message in $\{1, \ldots, 2^{n\hat{R}_1}\}$ to an index in $\{1, \ldots, 2^{nR_0^{(1)}}\}$.

viii     Also, randomly partition the message set $\{1, 2, \ldots, 2^{n\hat{R}_2}\}$ into $2^{nR_0^{(2)}}$ sets $\{S_1^{(2)}, S_2^{(2)}, \ldots, S_{2^{nR_0^{(2)}}}^{(2)}\}$ by independently and uniformly assigning each message in $\{1, \ldots, 2^{n\hat{R}_2}\}$ to an index in $\{1, \ldots, 2^{nR_0^{(2)}}\}$.

Encoding procedure Let $w_{1,k}$ and $w_{2,k}$ be the messages to be send on block $k$. $\mathcal{S}_1$ and $\mathcal{S}_2$ respectively transmit the codewords $x_1^n(w_{1,k})$ and $x_2^n(w_{2,k})$. We assume that $(u_1^n(s_{1,k-1}), \hat{y}_{r1}^n(z_{1,k-1}|s_{1,k-1}), y_r^n(k-1)) \in A_\epsilon^{(n)}$ and $z_{1,k-1} \in S_{s_{1,k}}^{(1)}$ and also that $(u_2^n(s_{2,k-1}), \hat{y}_{r2}^n(z_{2,k-1}|s_{2,k-1}), y_r^n(k-1)) \in A_\epsilon^{(n)}$ with $z_{2,k-1} \in S_{s_{2,k}}^{(2)}$. Then the relay transmits the codeword $x_r^n(s_{1,k}, s_{2,k})$.

Decoding procedure In what follows, we will only detail the decoding procedure at the receiver node $\mathcal{D}_1$ (at $\mathcal{D}_2$ the decoding is analogous). At the end of block $k$:

i     The receiver node $\mathcal{D}_1$ estimates $\hat{s}_{1,k} = s_1$ if and only if there exists a unique sequence $u_1^n(s_1)$ that is jointly typical with $y_1^n(k)$. We have $s_1 = s_{1,k}$ with arbitrarily low probability of error if $n$ is sufficiently large and $R_0^{(1)} < I(U_1; Y_1)$.

ii  Next, the receiver node $\mathcal{D}_1$ constructs a set $\mathcal{L}_1(y_1^n(k-1))$ of indexes $z_1$ such that $(u_1^n(\hat{s}_{1,k-1}), \hat{y}_{r1}^n(z_1|\hat{s}_{1,k-1}), y_1^n(k-1)) \in A_\epsilon^{(n)}$. $\mathcal{D}_1$ estimates $\hat{z}_{1,k-1}$ by doing the intersection of sets $\mathcal{L}_1(y_1^n(k-1))$ and $S_{\hat{s}_{1,k}}^{(1)}$. Similarly to [13, theorem 6] and using [13, lemma 3], one can show that $\hat{z}_{1,k-1} = z_{1,k-1}$ with arbitrarily low probability of error if $n$ is sufficiently large and $\hat{R}_1 < I(\hat{Y}_{r1}; Y_1|U_1) + R_0^{(1)}$.

iii  Using $\hat{y}_{r1}^n(\hat{z}_{1,k-1}|\hat{s}_{1,k-1})$ and $y_1^n(k-1)$, the receiver node $\mathcal{D}_1$ finally estimates the message $\hat{w}_{1,k-1} = w_1$ if and only if there exists a unique codeword $x_1^n(w_1)$ such that $(x_1^n(w_1), u_1^n(\hat{s}_{1,k-1}), y_1^n(i-1), \hat{y}_{r1}^n(\hat{z}_{1,k-1}|\hat{s}_{1,k-1})) \in A_\epsilon^{(n)}$. One can show that $w_1 = w_{1,k-1}$ with arbitrarily low probability of error if $n$ is sufficiently large and

$$R_1 < I\left(X_1; Y_1, \hat{Y}_{r1} \mid U_1\right). \tag{19}$$

iv  At the end of the block $k$, the relay looks for the suitable estimation of its observation that it intends to transmit to the receiver node $\mathcal{D}_1$ by estimating $\hat{z}_{1,k}$. It estimates $\hat{z}_{1,k} = z_1$ if there exists a sequence $\hat{y}_r^n(z_1|s_{1,k})$ such that $(u_1^n(s_{1,k}), \hat{y}_{r1}^n(z_1|s_{1,k}), y_r^n(k)) \in A_\epsilon^{(n)}$. There exists a such sequence if $n$ is sufficiently large and $\hat{R}_1 > I(\hat{Y}_{r1}; Y_r|U_1)$.

From i, ii, iii we further obtain

$$I(\hat{Y}_{r1}; Y_r|U_1, Y_1) < I(U_1; Y_1). \tag{20}$$

(19) and (20) are identical to the first terms in (5) and (6), respectively. The achievability proof for the second receiver node follows in a similar manner. Therefore, we have completed the proof.

## APPENDIX C
### ACHIEVABILITY PROOF OF THE RATE REGION IN THEOREM 5.3

**Details of the coding strategy**

Codebook generation

i  Generate $2^{nR_i}$ i.i.d. codewords $x_i^n(w_i) \sim \prod_{k=1}^n p(x_{i,k})$, where $w_i \in \{1, 2, \ldots, 2^{nR_i}\}$, $i \in \{1, 2\}$.

ii  Generate $2^{nR_0}$ i.i.d. codewords $x_r^n \sim \prod_{k=1}^n p(x_{r,k})$ and label them $x_r^n(s)$, $s \in \{1, 2^{nR_0}\}$.

iii  For each codeword $x_r^n(s)$, generate $2^{nR}$ conditionally independent codewords $\hat{y}_r^n \sim \prod_{k=1}^n p(\hat{y}_{r,k}|x_{r,k}(s))$ and label them $\hat{y}_r^n(z|s)$, $z \in \{1, \ldots, 2^{nR}\}$. For each pair $(x_r, \hat{y}_r) \in \mathcal{X}_r \times \hat{\mathcal{Y}}_r$, the conditional probability $p(\hat{y}_r|x_r)$ is defined as $p(\hat{y}_r|x_r) = \sum_{x_1,x_2,y_1,y_2,y_r} p(x_1) p(x_2) p(y_1, y_2, y_r|x_1, x_2, x_r) p(\hat{y}_r|y_r, x_r)$.

iv  Randomly partition the message set $\{1, 2, \ldots, 2^{nR}\}$ into $2^{nR_0}$ sets $\{S_1, S_2, \ldots, S_{2^{nR_0}}\}$ by independently and uniformly assigning each message integer in $\{1, \ldots, 2^{nR}\}$ to an index in $\{1, \ldots, 2^{nR_0}\}$.

Encoding procedure Let $w_{1,k}$ and $w_{2,k}$ be the messages to send in block $k$. $\mathcal{S}_1$ and $\mathcal{S}_2$ respectively transmit the codewords $x_1^n(w_{1,k})$ and $x_2^n(w_{2,k})$. We assume that $(x_r^n(s_{k-1}), \hat{y}_r^n(z_{k-1}|s_{k-1}), y_r^n(k-1)) \in A_\epsilon^{(n)}$ and $z_{k-1} \in S_{s_k}$. Then the relay transmits the codeword $x_r^n(s_k)$.

Decoding procedure The decoding procedures at the end of block $k$ are as it follows:

i  As a first step, the receiver node $D_i$, $i \in \{1, 2\}$, estimates $\hat{s}_k = s$ if and only if there exist a unique sequence $x_r^n(k)$ that is jointly typical with $y_i^n(k)$. We have $s = s_k$ with arbitrarily low probability of error if $n$ is sufficiently large and $R_0 < I(X_r; Y_i)$, $\forall i \in \{1, 2\}$, which we rewrite as $R_0 < \min\{I(X_r; Y_1), I(X_r; Y_2)\}$.

ii  Next, the receiver node $D_i$ constructs a set $\mathcal{L}_i(y_i^n(k-1))$ of indexes $z$ such that $(x_r^n(\hat{s}_{k-1}), \hat{y}_r^n(z|\hat{s}_{k-1}), y_i^n(k-1)) \in A_\epsilon^{(n)}$. $\mathcal{D}_i$ estimate $\hat{z}_{k-1}$ by doing the intersection of sets $\mathcal{L}_i(y_i^n(k-1))$ and $S_{\hat{s}_k}$. Similarly to [13, theorem 6] and using [13, lemma 3], one can show that $\hat{z}_{k-1} = z_{k-1}$ with arbitrarily low probability of error if $n$ is sufficiently large and $R < I(\hat{Y}_r; Y_i|X_r) + R_0$, $\forall i \in \{1, 2\}$, which can be written as $R < \min\left\{I\left(\hat{Y}_r; Y_1|X_r\right), I\left(\hat{Y}_r; Y_2|X_r\right)\right\} + R_0$.

iii  Using $\hat{y}_r^n(\hat{z}_{k-1}|\hat{s}_{k-1})$ and $y_i^n(k-1)$, $i \in \{1, 2\}$, the receiver node $\mathcal{D}_i$ finally estimates the message $\hat{w}_{i,k-1} = w_i$ if and only if there exists a unique codeword $x_i^n(w_i)$ such that $(x_i^n(w_i), x_r^n(\hat{s}_{k-1}), y_i^n(k-1), \hat{y}_r^n(\hat{z}_{k-1}|\hat{s}_{k-1})) \in A_\epsilon^{(n)}$. One can show that $w_i = w_{i,k-1}$ with arbitrarily low probability of error if $n$ is sufficiently large and

$$R_i < I\left(X_i; Y_i, \hat{Y}_r \mid X_r\right). \tag{21}$$

iv  At the end of the block $k$, the relay looks for the suitable estimation of its observation by estimating $\hat{z}_k$. It estimate $\hat{z}_k = z$ if there exists a sequence $\hat{y}_r^n(z|s_k)$ such that $(x_r^n(s_k), \hat{y}_r^n(z|s_k), y_r^n(k)) \in A_\epsilon^{(n)}$. There exists a such sequence if $n$ is sufficiently large and $R > I(\hat{Y}_r; Y_r|X_r)$.

From ii and iv we have that $I(\hat{Y}_r; Y_r|X_r) < \min\left\{I\left(\hat{Y}_r; Y_1|X_r\right), I\left(\hat{Y}_r; Y_2|X_r\right)\right\} + \min\{I(X_r; Y_1), I(X_r; Y_2)\}$ which is equivalent to

$$\max\left\{I\left(\hat{Y}_r; Y_r|X_r, Y_1\right), I\left(\hat{Y}_r; Y_r|X_r, Y_2\right)\right\} < \min\{I(X_r; Y_1), I(X_r; Y_2)\}, \tag{22}$$

because of the Markov chain $\hat{Y}_r \ominus (X_r, Y_r) \ominus Y_i$, $\forall i \in \{1, 2\}$. (21) and (22) are identical to (7) and (8), respectively.

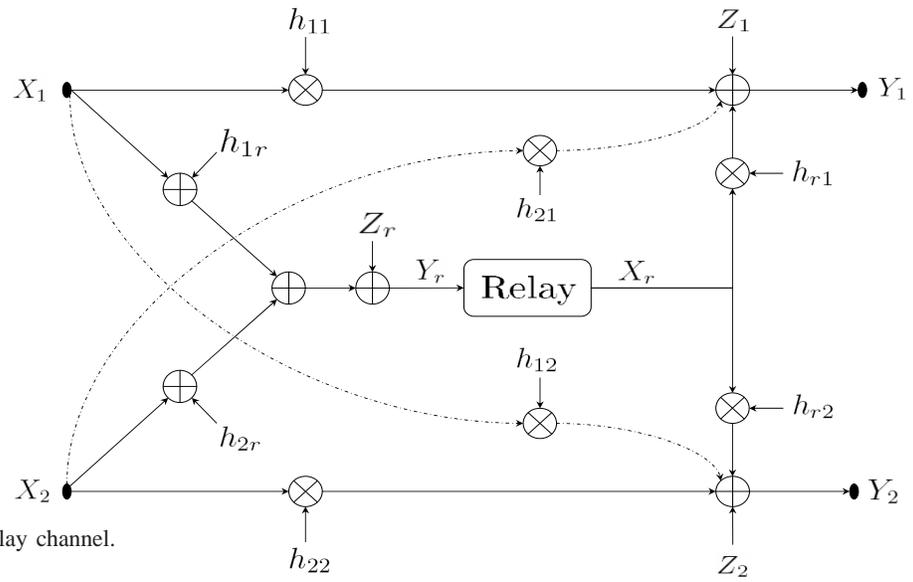

Fig. 1. Gaussian interference relay channel.

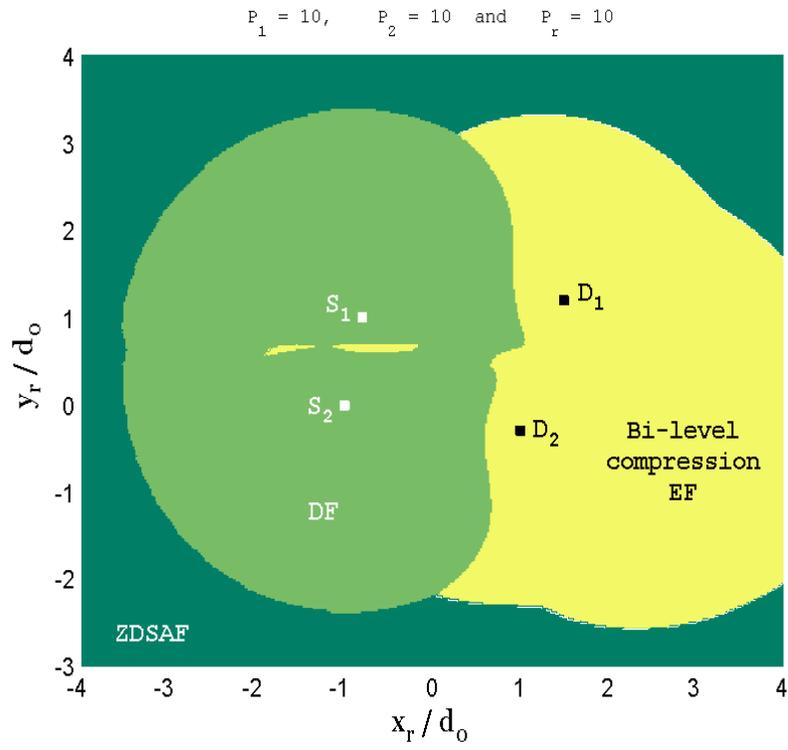

Fig. 2. For different relay positions in the plane $\left(\frac{x_r}{d_0}, \frac{y_r}{d_0}\right) \in [-4, +4] \times [-3, +4]$, the figure indicates the regions where one relaying protocol (AF, DF or bi-level EF) dominates the two others in terms of network sum-rate.

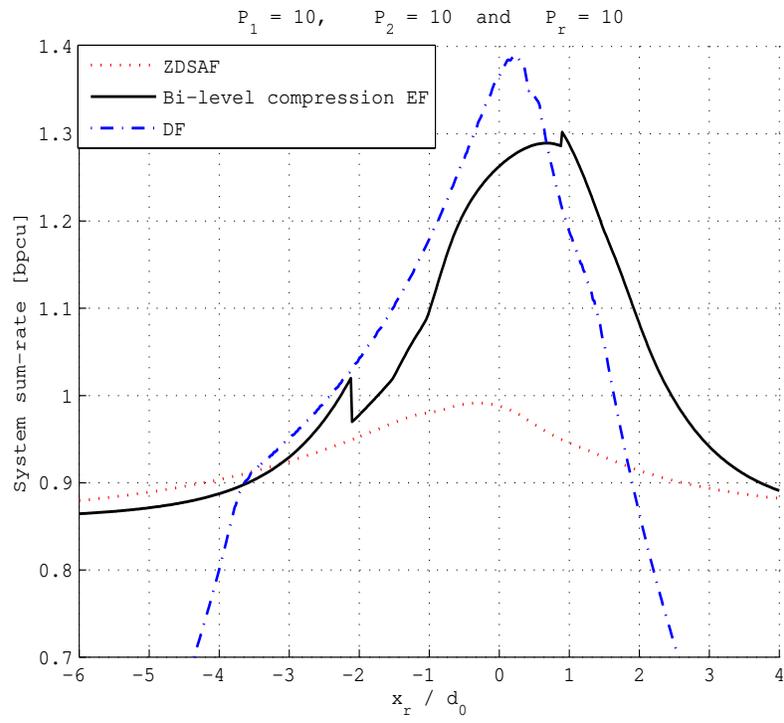

Fig. 3. Achievable system sum-rate versus $x_r$ (abscissa for the relay position) for a fixed $y_r$ ($y_r = 0.5d_0$), with AF, DF and bi-level EF.

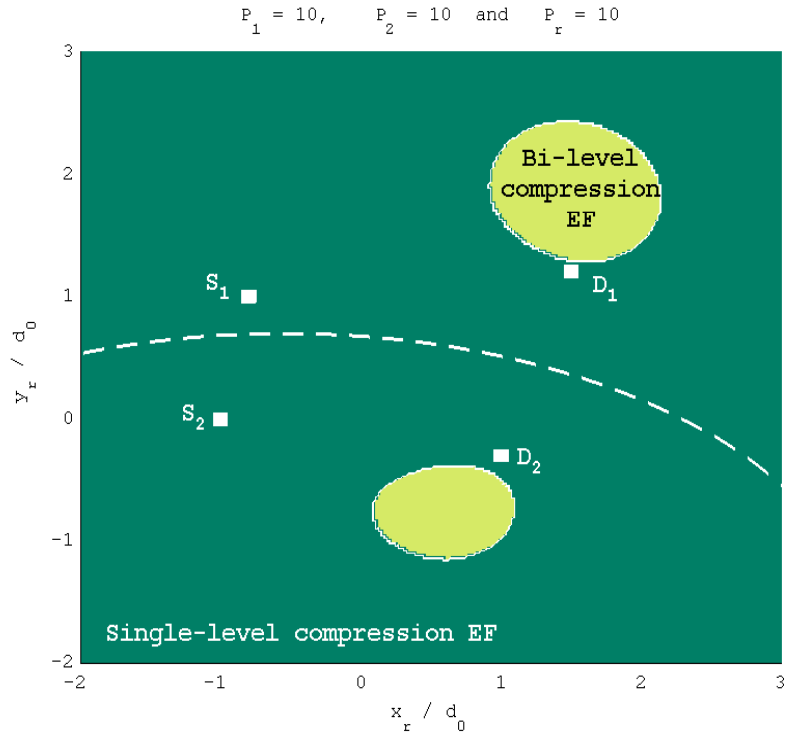

Fig. 4. Symmetric scenario in terms of transmit power: for different relay positions in the plane $\left(\frac{x_r}{d_0}, \frac{y_r}{d_0}\right) \in [-2, +3] \times [-2, +3]$, the figure indicates the regions (around $\mathcal{D}_i$) where the bi-level compression EF protocol performs better than its single-level counterpart in terms of network sum-rate.

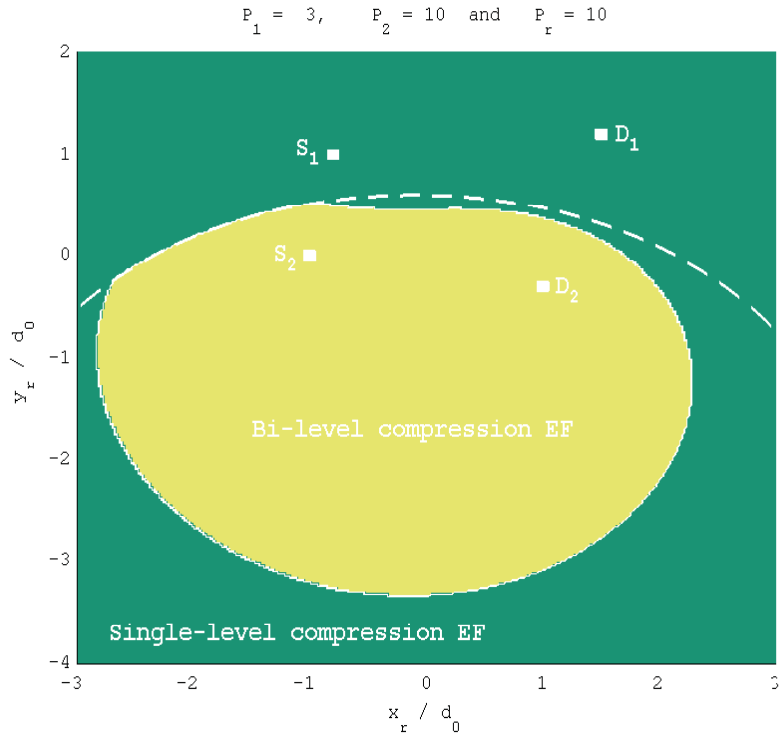

Fig. 5. Asymmetric scenario in terms of transmit power: for different relay positions in the plane $\left(\frac{x_r}{d_0}, \frac{y_r}{d_0}\right) \in [-2, +3] \times [-2, +3]$, the figure indicates the regions (around $\mathcal{D}_i$) where the bi-level compression EF protocol performs better than its single-level counterpart in terms of network sum-rate.